# Spontaneous growth of perfectly circular domains of MoS$_2$ monolayers using chemical vapour deposition technique

Umakanta Patra, Subhabrata Dhar*

Department of physics, Indian institute of technology, Bombay, India, 400076

Corresponding author email: dhar@phy.iitb.ac.in

**Abstract:**

Very large-scale integration of devices in a circular pattern has several advantages over the commonly used rectangular grid layout. For the development of such integrated circuits on a 2D semiconductor platform, spontaneous growth of the material in the form of circular islands is desirable. Here, we report the natural formation of 1L-MoS$_2$ circular islands of diameter as large as a few hundreds of micrometer on SiO$_2$/Si substrates by chemical vapor deposition (CVD) technique without the use of any seeding layer. The size of the circles is found to increase with the amount of sulphur used during growth. The study reveals that these circular islands are formed with a less-defective interior and a more-defective outer part that is dominated by a large density of grain boundaries and twists. Due to the lower defect density, the interior region yields much higher photoluminescence than the peripheral part. Field effect transistors (FETs) are fabricated on inner and outer portions of a circle to estimate the mobility and concentration of the background carriers in the two regions. The study shows that the maximum mobility is more than double in the interior than the outer part. While the carrier concentration remains practically unchanged in the two regions. The natural tendency to minimize the strain energy resulting from the mismatch between the thermal expansion coefficients of the monolayer and the substrate as well as the edge energy that originates from the boundary tension are thought to be the driving forces behind the formation of these circular domains.

## I. INTRODUCTION

Two-dimensional (2D) semiconductors, such as MoS$_2$, WS$_2$, MoSe$_2$ and WSe$_2$, which belong to transition metal dichalcogenides (TMDs) family, have gained a lot of attentions with MoS$_2$ being the most studied material [1–6]. Monolayer (1L) TMDC with direct band gap [7], high excitonic binding energy (~500 meV) [8], inversion asymmetry and strong spin-orbital coupling have great prospects for the development of future flexible and transparent electronics [9], valleytronics [10] and excitonics [11]. 2D TMD layers are often obtained by mechanical exfoliation technique [12]. Even though the density of defects in exfoliated flakes is found to be much less than those grown by other techniques, the method does not offer much control on the layer number, shape and size [13]. Chemical vapour deposition (CVD) technique is a promising method to grow large area covered films [6]. CVD gives a better control over the layer thickness and to some extent on the shape and size of the grown layers [1]. Atomically thin layers of TMDC materials with high stiffness, breaking strength and larger band gap than silicon [14] are considered to be superior choice for the suppression of the short channel effects to achieve larger density of devices in a chip [15]. Controlling the shape and size of the 2D layers is essential to fully utilize them for future electronic applications [16,17]. MoS$_2$ layers of various geometrical shapes, such as triangles [18], hexagons [19], dendritic [20], three and six-point stars [21,22] etc are reported in the literature. All these shapes are consistent with the inherent triangular symmetry of the lattice. It should be noted that one of the methodologies in designing very large-scale integrated (VLSI) circuits is to arrange the transistors and the interconnects in circular patterns rather than the traditional rectangular grid layout [23]. Such a design offers several advantages like reduced parasitic capacitances/resistances, better area utilization, improved noise rejection, enhanced thermal distribution, improved matching of differential pairs and shorter interconnect length. In this context, it will be interesting if these 2D layers can be spontaneously grown in the form of circles. There are in fact a few reports on the growth of round shape structures of 2D layers [24–26]. Shuang Xie *et al.* have shown that transformation from triangular to round-shaped domains of 1L-MoS$_2$ on SiO$_2$/Si substrates is possible in CVD technique by adjusting the growth temperature and the separation between the substrate and the MoO$_3$ precursor [25]. B. Ryu et al. has used array of platinum (Pt) mesas as seeding layers to induce heterogeneous nucleation of MoS$_2$ on SiO$_2$/Si substrates in CVD technique [24]. These mesas are found to facilitate the site selective growth of rounded 1L-MoS$_2$ islands surrounding them. Formation of arrays of such structures, each of which is a few tens of micrometer in diameter, has been demonstrated by this technique.

Here, we report the spontaneous formation of 1L-MoS$_2$ circular domains with diameter as large as hundreds of micrometer on SiO$_2$/Si substrates using CVD technique without introducing any seeding layer. Moreover, it has been observed that the diameter of the circles increases with the amount of sulphur powder used for the growth. The study reveals that the outer region of the circles are dominated by a large density of grain boundaries and twists. While the inner part contains much lower density of such defects. This nonuniformity of defect distribution results in much higher PL yield from the central region as compared to the peripheral part of the circular domains. Back-gated FET structures are fabricated on the inner and outer portions of a circle and the carrier mobility is estimated from the transconductance profiles of these devices. The maximum mobility value has been found to be 2.5 times higher in the inner region than the outer portion. These circular domains with the nonuniform distribution of grain boundaries and twists are believed to be resulting from the natural tendency to minimize two energies, namely the strain energy, which originates from the mismatch between the thermal expansion coefficients of the monolayer and the substrate and the edge energy that arises from the boundary tension.

## II. METHODS

Samples were grown on SiO$_2$ coated Si substrates by chemical vapour deposition technique. A schematic diagram of the reactor is presented in figure 1(a). High purity (5N) MoO$_3$ and sulphur (S) powders were used as precursors. Placement of the precursor containing boats and the substrate are also shown in the figure. Substrate was placed on the floor of the MoO$_3$ containing alumina boat. Substrates were cleaned successively in trichloroethylene (TCE), acetone and propyl alcohol (IPA) solutions for 10 min each under ultrasonication before drying in the flow of N$_2$ gas. Several samples were grown. In all cases, MoO$_3$ amount was fixed at 6 mg, while the amount of sulphur (S) powder was varied from 350 to 450 mg. Flow rate of argon, which had been used as the carrier gas, was maintained at 7 sccm during the growth. Samples were grown at 700 ℃.

Optical images were recorded using an Olympus (BX53-F) microscope. Atomic force microscopy (AFM) was carried out in tapping mode. A micro-PL/Raman setup (Renishaw, Invia Reflex, UK) was used for photoluminescence (PL) and Raman study. A 532 nm diode laser with a power of 500 $\mu$W was used for excitation. PL and Raman spectra were obtained using a 500 cm focal-length monochromator equipped with 2400 gr/mm grating and CCD detector. PL maps were recorded with a 50× magnification objective lens and scanning step size of 1 $\mu$m. High resolution transmission electron microscopy (HRTEM) of these samples was carried out in a 300 kV system taken from Thermo Scientific (Themis 300 G3) after transferring the films onto a Cu-grid using a polystyrene assisted wet transfer technique [11].

MoS$_2$ islands were transferred onto a 300 nm thick SiO$_2$ coated p-type Si-substrates using the polystyrene assisted transfer technique. Backgated field effect transistor (FET) devices were fabricated using standard optical lithography and lift-off technique. Ni (30nm) and Au (100 nm) contacts were deposited using sputtering technique. The current-voltage characteristics of the FETs were recorded using Polaris probe station (Keysight B1500 (Agilent) SDA/Karl Suss/ PM8).

## III. RESULTS AND DISCUSSION

Figure 1(b-d) show the optical images of the circular islands in samples grown with different amounts of sulphur powder. Evidently, the diameter of the largest circle is about 250 $\mu$m, which is observed in the sample grown with 450 mg of sulphur. It has been found that the average size of the circles increases with the sulphur amount used during growth. This can be seen in the supplementary figure S1. Figure 1(e) compares the room temperature Raman spectra recorded at points lying in the inner and outer parts of a circular island for the sample grown with 350 mg of sulphur powder. Both the spectra, which are featured by the characteristics peaks associated with the $E_{2g}$(in-plane) and $A_{1g}$ (out of-plane) vibrational modes of MoS$_2$, almost overlap onto each other. Peak separation between $E_{2g}$ and $A_{1g}$ features is 19.5 cm$^{-1}$. These findings strongly suggest that the circular islands are the monolayers of MoS$_2$ [5]. Atomic force microscopy is carried out to examine whether the circles are indeed monolayers or not. Figure 1(f) shows the AFM image taken at the edge of a circle. The boundary is clearly visible. Inset shows the height profile recorded along the white line as marked in figure 1(f). The estimated thickness of 0.85 nm matches very well with the expected height of a monolayer MoS$_2$. [27] Similar studies are carried out on several circles across the grown layers, which confirm monolayer nature of these islands. In figure 1(b-d), one can notice a dark spot at the centre of every circular island. These are likely to be the residues of the precursor materials, which act as the nucleation centres. This finding might imply that the seeding of these circular layers starts at the centre.

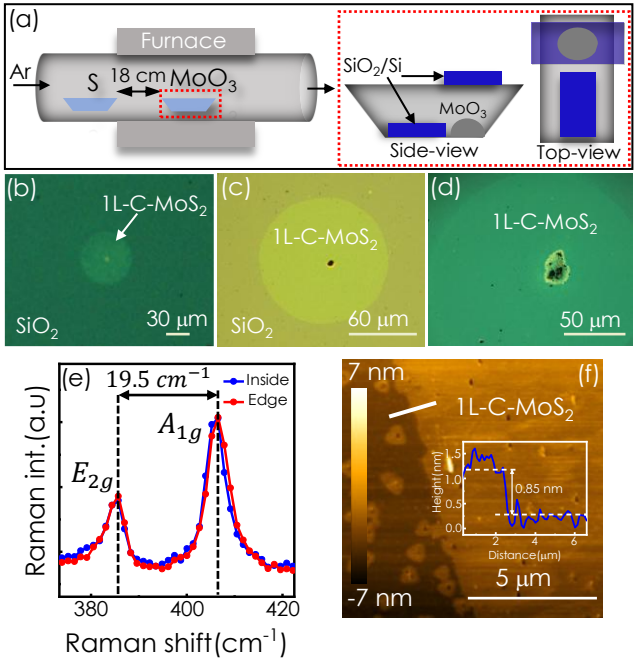

**FIG. 1.** (a) Schematic depiction of the CVD set-up. Optical images of the circular MoS$_2$ islands in samples grown with sulphur (S) amount of (b) 350 mg, (c) 400 mg and (d) 450 mg. (e) Room temperature Raman spectra recorded from inner (Blue) and outer (Red) regions of a circular island. (f) AFM image showing the edge of the circular MoS$_2$. Inset: height scan profile recorded along the white line marked in the image.

Figure 2(a) and (b) show the AFM images recorded from the central and peripheral regions, respectively. There are certain interesting differences in the two images. In both the images, a large number of triangular shaped dark spots are visible on the island surface. However, their density in the outer region is more than the inner region. Moreover, the peripheral zone also has a large density of smaller spots, which coexist with certain filament-like structures. Few of these filaments are marked by arrows. From the shape and the contrast, the bigger spots appear to be secondary growth of multilayer MoS$_2$. We believe that the smaller spots are also secondary MoS$_2$ growth. It is plausible that these small nucleation centres are forming preferentially at the grain boundaries. Filament-like structures might be forming due to the coalescence of these nucleations along the grain boundaries. This picture gains strength as one can find linear chain of spots on various locations in the peripheral region of the circle [marked by blue boundaries in Fig. 2(a)]. These spots are likely to be the nucleation centres, which are formed along a grain boundary at such a low density that they cannot coalesce to form a filament. We have also recorded TEM images from different parts of a circular island after transferring it on a Cu-grid [11]. In the boundary region, secondary nucleations as well as the formation of grain boundaries and filaments can be observed [see supplementary Fig. S2]. All these findings indicate that the outer region of the circular island is dominated by grain boundaries and secondary growths. Fig. 2(c) presents the room temperature PL intensity map for one of the circular islands. The central region of the circle is clearly much brighter than the peripheral part, which has a grainy dark-bright contrast. Figure 2(d) shows the room temperature PL spectra recorded from the inner and outer regions of the circular island. Interestingly, the PL spectrum obtained from the outer region peaks at a slightly lower energy than that is recorded from the central part [peak energy 1.87 eV]. This is more clearly visible in the inset of the figure. Note that 1.87 eV is the PL peak energy typically reported for 1L-MoS$_2$ at room temperature. Lower intensity and the red-shift of the PL feature can be attributed to the large density of secondary nucleation centres found in the outer regions of the circular islands.

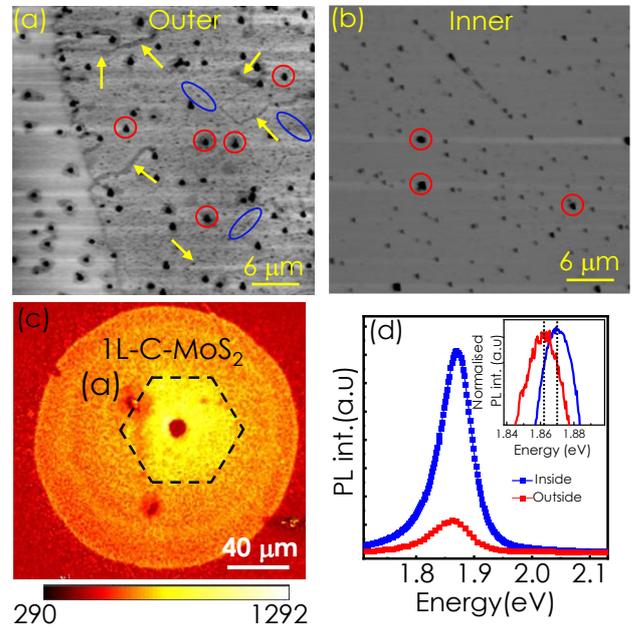

**FIG. 2.** AFM images recorded from the (a) central and (b) peripheral regions of a circular island. (c) Room temperature PL intensity map of the island. (d) Room temperature PL spectra recorded from spots lying in the inner (blue) and outer (red) regions of the circle. Inset shows the PL spectra around the peak positions in an expanded scale.

Figure 3(a) and (b) show the high-resolution TEM images recorded from the inner and outer regions, respectively, of a circular island after transferring it on a Cu-grid [11]. In both the images hexagonal arrangements of the atoms are clearly evident. Selected area electron diffraction (SAED) patterns recorded for an inner and an outer spot of the island are shown in panels (c) and (d), respectively. SAED pattern [Fig.3(c)] from the point lying in the inner part of the circle shows hexagonally arranged spots as expected for a MoS$_2$ single domain. While the SAED recorded from the outer zone [Fig.3(d)] presents an overlapping pattern of two hexagonal arrangements rotated by certain angle suggesting the existence of rotated grains.

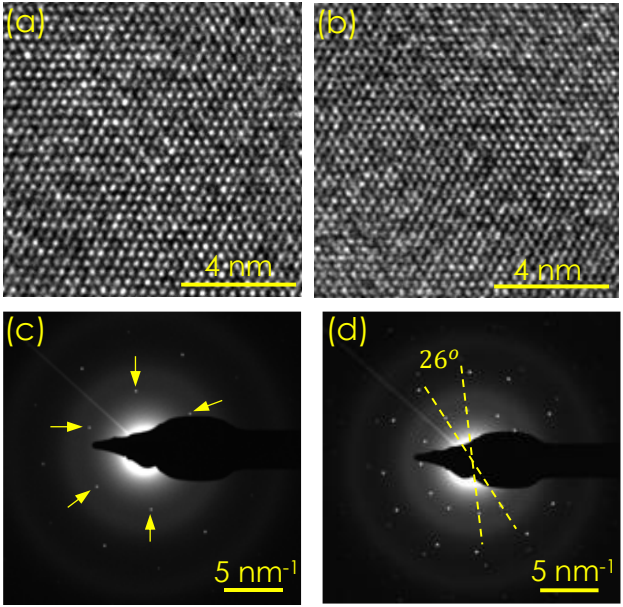

**FIG. 3.** High resolution TEM images recorded from an (a) inner and (b) outer spots of a circular island after transferring it to a Cu-grid. Selected area electron diffraction (SAED) patterns recorded from an (c) inner and a (d) outer spot of the island.

We have analysed several SAED patterns recorded randomly from different locations on the circular island [see supplementary Figure S3]. While the patterns recorded from the interior parts mostly represent single domain of MoS$_2$, the data from the outer areas primarily show overlapping patterns with different angles of rotations. These findings suggest the presence of high density of misoriented domains in the outer region of the island, which is consistent with the observations of Fig.3 and supplementary FIG. S2.

Field effect transistors are fabricated on a monolayer circular MoS$_2$ (grown with 400 mg of Sulphur precursor) after transferring it onto a fresh SiO$_2$/Si substrate. Note that the intensity variation and the shift of the PL spectra recorded from the inner and outer regions follow the same trend before and after the transfer [see supplementary figure S4]. Electrical properties of the inner and outer regions of the circular island have been studied using the back gated FET devices fabricated on both the regions as shown schematically in Figure 4(a). Figure 4(b) plots the variation of source-drain current ($I_{ds}$) with the source-drain voltage ($V_{ds}$) at zero source-gate bias ($V_{gs}$). The conductivity of the inner region is clearly more than the outer region. Figure 4(c) presents the $I_{ds}$ versus $V_{gs}$ profiles of the inner and outer FETs recorded at $V_{ds} = 10\ V$. $I_{ds}$ is found to be 0.50 μA at $V_{gs} = 30\ V$ for inner region, however $I_{ds}$ reduce to 0.31 μA for outer region. $I_{ds}$ is found to be 0.50 μA at $V_{gs} = 30\ V$ for inner region, however $I_{ds}$ reduce to 0.31 μA for outer region. The transconductance $g_m = dI_{ds}/dV_{gs}$ has been plotted as a function of $V_{gs}$ in the inset of the figure. The maximum value of $g_m$ is clearly more in the inner region than the outer. The maximum carrier mobility $\mu$, which follows the relation $\mu = g_m L/C_g W V_{ds}$ [28] with $C_g$ is the gate capacitance (11.5 nFcm$^{-2}$ for 300 nm thick SiO$_2$ gate), $L$ (~5 μm) is the channel length and $W$ (~7 μm) is the channel width, can be estimated as 0.1 and 0.25 cm$^2$/V-s, respectively, for the outer and the inner channels.

Threshold voltages ($V_{th}$) for the two channels have been estimated from the $\sqrt{I_{ds}}$ versus $V_{gs}$ plots shown in figure (d). $V_{th}$ is clearly more negative for the inner than the outer channel, which suggests slightly higher background electron concentration in the former. Background electron concentration ($n$) has been estimated using the relation $n = C_g(-V_{th})/e$, [29] where $e$ is the electronic charge, as $1 \times 10^{12}$ and $1.4 \times 10^{12}$ cm$^{-2}$ for the outer and inner regions, respectively. These findings indicate that the main reason for higher conductivity in the inner region is 2.5 times enhancement of electron mobility there as compared to the outer region. Lower mobility in the outer region can be attributed to higher density of grain boundaries, which are expected to contribute to the carrier scattering.

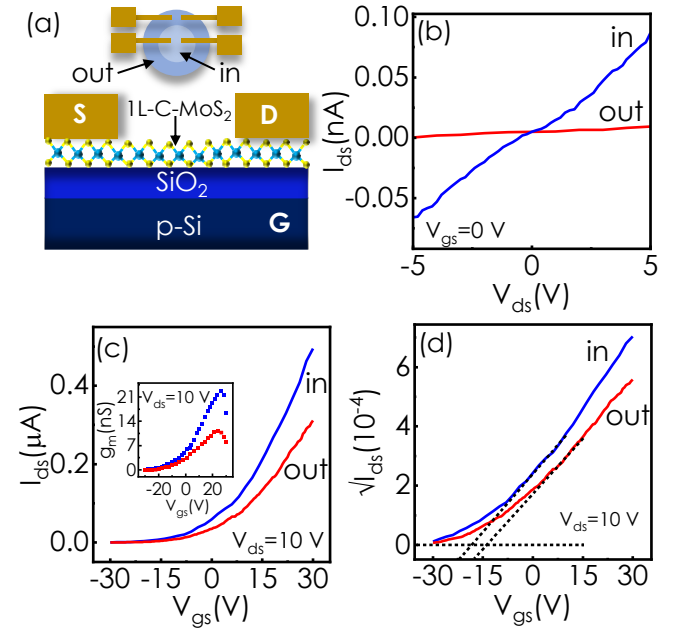

**FIG. 4.** (a) Schematic view of the FET device structure. Inset: source and drain contacts for the inner and outer channels. (b) $I_{ds}$ versus $V_{ds}$ profiles recorded at zero gate bias, (c) $I_{ds}$ verses $V_{gs}$ profiles recorded at $V_{ds} = 10\ V$ and (d) Square root of $I_{ds}$ verses $V_{gs}$ plots for the inner and outer channels. Inset of (c):

transcondutance ($g_m = dI_{ds}/dV_{gs}$) as a function of $V_{gs}$ for the two channels.

The question of how such circular islands of 1L-MoS$_2$ are formed, when the crystalline symmetry of the material is hexagonal is really interesting. We believe that at the growth temperature, single-crystal hexagonal islands are first formed. Existence of a spot at the centre of every circle suggests a common nucleation point for the entire island. However, as the temperature is reduced after the growth, the strain energy originating from the mismatch between the thermal expansion coefficients of the monolayer and the substrates increases. Edge energy that arises from the boundary tension also becomes important. Note that the edge energy per unit area of a hexagonal island is more than that of a circle. The two energies play the crucial role in governing the shape and morphology of the island. While the strain energy can be reduced through creation of grain boundaries, the edge-energy minimization demands a transformation of shape from hexagon to circle. It is plausible that the minimization of both the energies results in the crack generation in the island in such a way that leaves a radial distribution of grain size with bigger grains at the central region and smaller ones at the outer region as schematically depicted in Fig. 5. The shape change of the island takes place via two pathways; grain-detachment and grain-rotation. The grains at the corners of the hexagonal island are detached to round-off the edges. Note that the presence of such detached grains could indeed be observed at the boundary of such a circular island in TEM [see supplementary figure S2(a)]. Grains lying at the outer region of the main island also rotates to accommodate the shape change. Note that such rotations are indeed observed in Fig. 3.

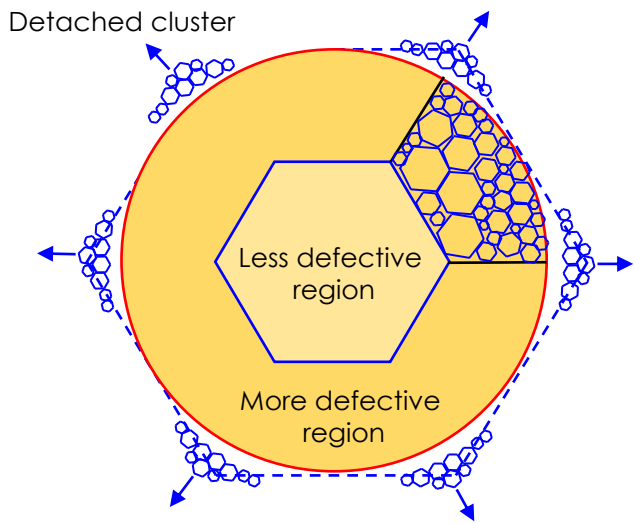

**FIG. 5.** Schematic depiction of the formation mechanism of the circular 1L-MoS$_2$ islands.

## IV. CONCLUSION

Circular domains of 1L-MoS$_2$ with diameter as large as several hundreds of micrometres can be formed spontaneously on SiO$_2$/Si substrates by CVD technique without using any seeding layer. Average size of the circular domains increases with the amount of S-powder used during growth. These circles are found to be consisting of highly defective outer regions dominated by grain boundaries and twists and relatively defect free central regions. Lower density of defects results in the higher PL yield and carrier mobility in the central parts as compared to the outer regions of these circles. Natural tendency to minimize the strain energy that results from the mismatch between the thermal expansion coefficients of the monolayer and the substrates as well as the edge energy arising from the boundary tension leads to the formation of these circular domains.

**ACKNOWLDGEMENT**

We acknowledge the Science and Engineering Research Board (SERB), under the grant number CRG/2022/001852, Government of India for the financial support. We also thankful to Sophisticated Analytical Instrument Facility (SAIF), Industrial Research and consultancy Centre (IRCC), and the Centre for Excellence in Nanoelectronics (CEN), IIT Bombay for providing the characterization facilities.

# Supplementary materials for

# Spontaneous growth of perfectly circular domains of MoS$_2$ monolayers using chemical vapour deposition technique

Umakanta Patra, Subhabrata Dhar*

Department of physics, Indian institute of technology, Bombay, India, 400076.

Corresponding author email: dhar@phy.iitb.ac.in

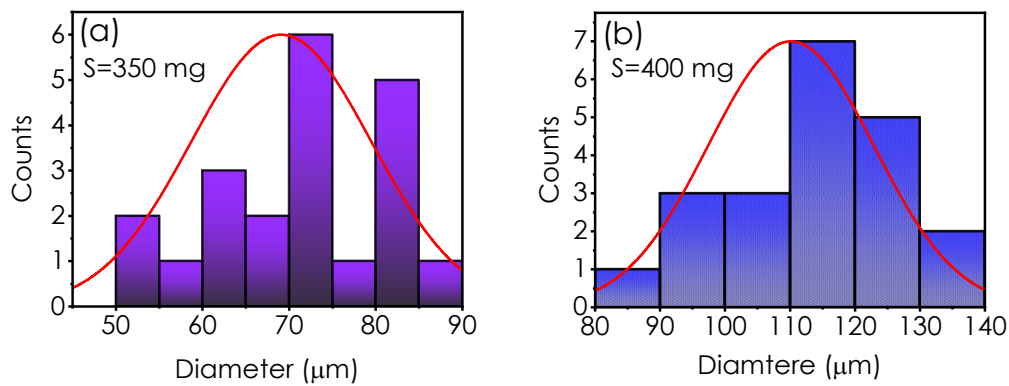

**FIG. S1.** Histogram plots for the diameter distribution of the 1L-MoS$_2$ circles in samples grown with (a) 350 mg (b) 400 mg of sulphur. Average size is found to be ∼ 70 and ∼ 110 μm, respectively, in samples grown with sulphur amounts of 350 and 400 mg. Note that for the sample grown with 450 mg of sulphur, very few but larger size (diameter > 200 $\mu$m) 1L-MoS$_2$ circles could be seen.

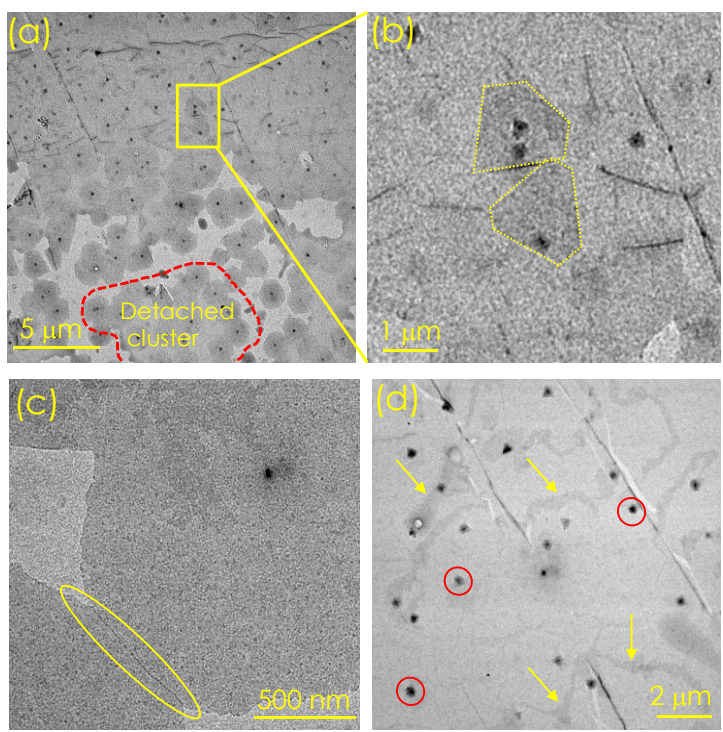

**FIG. S2.** Low magnification HRTEM image recorded at the outside region of a circular 1L-MoS$_2$ island. (a) Several detached domains and their clusters could be seen at the boundary of the circular island. (b) Magnified view of the portion bounded by the yellow box in (a). Two rotated domains could be seen very clearly there [marked by yellow dotted lines]. (c) Another portion of the outer region, where the grain boundary is clearly visible. (d) Another portion of the outer region, where the secondary growth of filament-like structures is quite evident.

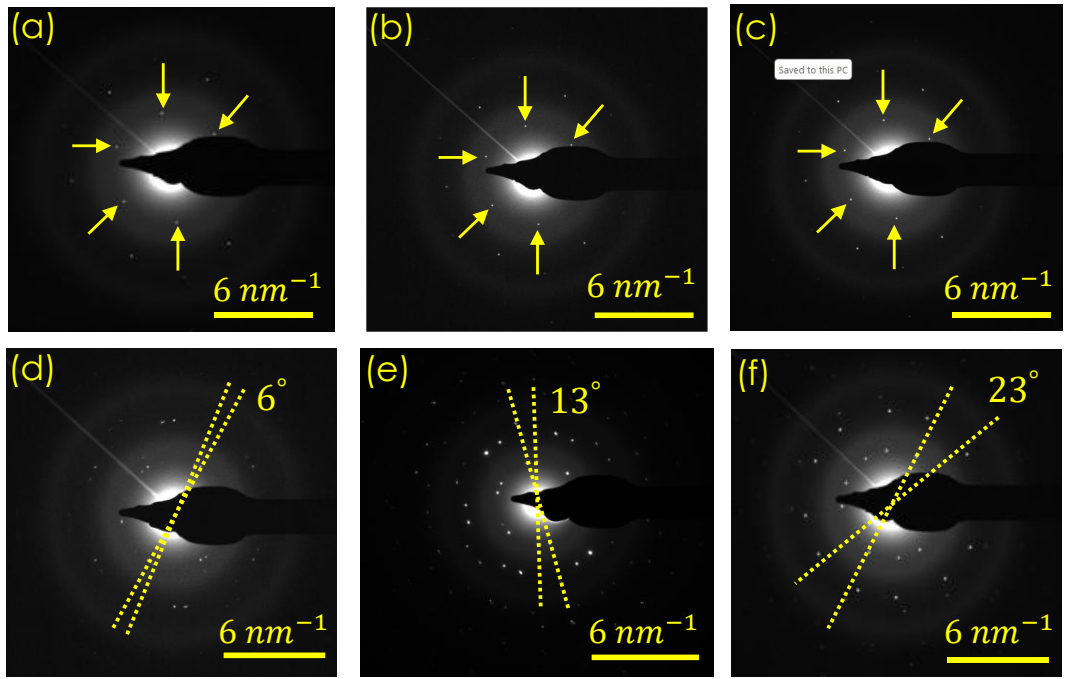

**FIG. S3.** Selected area electron diffraction (SAED) patterns recorded from the inner regions (a-c) and outer regions (d-f) of a circular 1L-MoS$_2$ island. Presence of hexagonally arranged diffractions sports are evident in the inner regions. While in the outer regions, double spotted diffraction patterns are visible. The double-spot formation suggests the rotation of the grains. Various angles of grain rotation could be seen in these patterns.

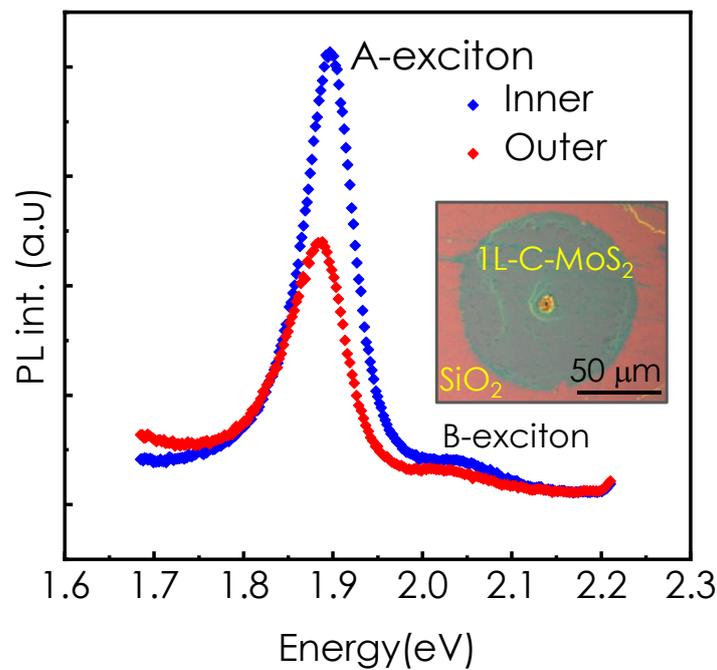

**FIG. S4.** Compares the PL intensity recorded from the inner and outer regions of a circular 1L-MoS$_2$ after transferring it onto a SiO$_2$/Si substrate. Inset shows the optical image of the island after the transfer.